\documentclass{article}%
\usepackage[open,openlevel=1]{bookmark}
\usepackage[most]{tcolorbox}
\usepackage{alphabeta}
\usepackage{amsmath}
\usepackage{amsfonts}
\usepackage{amssymb}
\usepackage{graphicx}
\usepackage{boxedminipage}
\usepackage{multirow}
\usepackage{algorithmicx}
\usepackage{algpseudocode}
\usepackage{caption}
\usepackage{listings}
\usepackage{color}%
\providecommand{\U}[1]{\protect\rule{.1in}{.1in}}
\newtheorem{theorem}{Theorem}[section]

\newtheorem{definition}[theorem]{Definition}
\newtheorem{example}[theorem]{Example}

\newtheorem{problem}[theorem]{Problem}

\begin{document}

\title{An Algorithm for \\Limited Visibility Graph Searching}
\author{Ath. Kehagias and A.C. Papazoglou}
\date{\today}
\maketitle

\begin{abstract}
We study a \emph{graph search }problem in which a team of \emph{searchers}
attempts to find a \emph{mobile target} located in a graph. Assuming
that\ (a)\ the \emph{visibility field} of the searchers is limited, (b)\ the
searchers have unit speed and (c)\ the target has infinite speed, we formulate
the \emph{Limited Visibility Graph Search} (LVGS)\ problem and present the
LVGS\ algorithm, which produces a \emph{search schedule} guaranteed to find
the target in the minimum possible number of steps. Our LVGS algorithm is a
\textquotedblleft conversion\textquotedblright\ of Guibas and Lavalle's
\emph{polygonal region} search algorithm .

\end{abstract}

\setcounter{tocdepth}{4}

\section{Introduction\label{sec01}}

In \emph{graph search }problems a team of \emph{searchers} attempts to find a
\emph{mobile target} located in a graph. Depending on the properties of the
searchers and target, different variants of the problem are obtained.

In the current paper we study a graph searching variant in which it is assumed
that: (a)\ the \emph{visibility field} of the searchers is limited (i.e., from
any given position they can see only part of the graph, as described by a
\emph{visibility matrix}),\ (b)\ the searchers have unit speed and (c)\ the
target has infinite speed\footnote{These terms will be defined precisely in a
later section.}. We formulate, under these assumptions, the \emph{Limited
Visibility Graph Search} (LVGS)\ problem and present an LVGS\ algorithm which,
given the adjacency and visibility matrices, produces a \emph{search schedule}
guaranteed to find the target (if this is possible) in the minimum possible
number of steps. Our LVGS algorithm is a \textquotedblleft
conversion\textquotedblright\ of the \emph{polygonal region} search algorithm
presented in \cite{Guibas1997,Guibas1999}.

The seminal papers for \emph{graph search problems} are \cite{Breisch1967} and
\cite{Parsons1978}. The basic ideas of these two works have been elaborated
and extended in a large number of papers\footnote{It is worth emphasizing that
in this formulation of the problem a worst-case analysis is adopted, by which
the target is essentially \textquotedblleft factored out\textquotedblright\ of
the problem; i.e., the only decision maker is the searcher. This point will be
further explained in later sections.}; excellent reviews of the relevant
literature appear in \cite{Fomin2008} and \cite{Hollinger2011}. A large number
of graph search algorithms have been presented in the literature; see for
example
\cite{Alspach2006,Barriere2002,Dendris1997,Kiroussis1986,Seymour1993,Yang2007}%
. These algorithms usually depend on path- or tree-decompositions of the graph
(and related methods) resulting in rather complicated implementation; in
addition, the majority of the above works concerns the \emph{zero}-visibility
case, i.e., when the searchers can only see the vertex in which they are
located. Furthermore, the search schedules studied in the graph theoretic
literature often allow a searcher to move between non-adjacent vertices (i.e.,
not necessarily following the edges of the graph). This movement mode is
called \textquotedblleft teleporting\textquotedblright\ and is not appropriate
for most realistic applications. Relatively little has been published
\cite{Barriere2002,Barriere2003,Deren2011,Deren2012} on non-teleporting
search, i.e., where the searchers move only along the graph edges. Some
related approaches appear in the robotics literature, for example in
\cite{Hollinger2009,Hollinger2010,Hollinger2010b,Sarmiento2004}, where it is
usually assumed that the target is invisible but has finite speed.

A related, but different, approach to limited visibility graph searching can
be obtained in the context of \emph{pursuit games in graphs\footnote{In this
formulation we are dealing with a \emph{game} in which both the pursuers and
the evader are active decision makers. }}. These involve one or more
\emph{pursuers} (i.e., searchers)\ who try to locate \emph{and capture} an
\emph{evader} (i.e., target); all agents are assumed to move along the edges
of the graph (and usually, but not always, with unit speed). The
\textquotedblleft classic\textquotedblright\ version of this problem is the
\emph{Cops and Robbers Game} (CR) first introduced in
\cite{Nowakowski1983,Quilliot1985}, in which it is assumed that the cops
(searchers) are always aware of the robber's (target's) location; so this is a
\emph{complete} visibility problem. Many CR\ variants have been considered,
obtained by varying the abilities and restrictions placed on both the cops and
the robber. There is a large literature on the Cops and \emph{visibile }
Robber problem, mainly concentrating on its theoretical aspects. An excellent
and recent review of the available work appears in the book
\cite{NowakowskiBook}. Algorithmic aspects of the problem are studied in
\cite{Berarducci1993} and \cite{Hahn2006}.

The variant of an \emph{invisible} or \emph{partially visible} robber is quite
close to the previously mentioned LVGS\ problem. The cops (resp. the
robber)\ play a role similar to that of the searchers (resp. target); note,
however, the distinction between \emph{seeing} and \emph{capturing }the
robber. Relatively little\ work has been done on this problem. The first
formulation appears in \cite{Tosic1985}. Further progress was achieved in the
Master's theses \cite{Jeliazkova2006,Tang2004} and, recently, in
\cite{Deren2013,Deren2015,Deren2015b} (for the zero-visibility case) and in
\cite{Clarke2019} (for the general case, which includes zero- and
complete-visibility as special cases). A somewhat different approach to the
invisible robber appears in \cite{Kehagias2012,Kehagias2014} and a related
computational approach is presented in \cite{Kehagias2009}.

The approach we follow in the current paper is rather different from all of
the above. As already mentioned, our formulation and solution of the problem
are based on the ideas of \cite{Guibas1997,Guibas1999}, which studies search
of a \emph{continuous polygonal environment}. This approach and
generalizations to various geometric environments is presented more
extensively in the excellent book \cite{Lavalle2006}.

In contrast to the previously mentioned treatments of graph search and
pursuit, our LVGS algorithm is relatively simple to implement, produces
non-teleporting search schedules and can be applied not only to
zero-visibility but to \emph{arbitrary} visibility specifications. On the
other hand, while the algorithm can be applied to any number $K$ of searchers,
the computational burden increases quickly with $K$. In short, the main
usefulness of the LVGS\ algorithm is in solving the graph search problem under
the following conditions.

\begin{enumerate}
\item \emph{One} searcher is attempting to locate the target in a graph.

\item Both searcher and target move only along the edges of the graph; the
searcher moves with unit speed while the target moves with infinite speed.

\item The \emph{visibility} of the target is determined by the visibility
matrix $B$.
\end{enumerate}

The paper is organized as follows. In Section \ref{sec02} we introduce
preliminary definitions and notations. In Section \ref{sec03} we formulate the
LVGS\ problem and present the LVGS algorithm. Section \ref{sec04} is devoted
to the evaluation of the algorithm by numerical experiments. Finally, in
Section \ref{sec05} \ we summarize our work and propose directions for further research.

\section{Preliminaries\label{sec02}}

We use the following set theoretic notation.

\begin{enumerate}
\item Given a set $U$, the \emph{cardinality} of $U$ is denoted by $\left\vert
U\right\vert $ and defined to be the number of elements of $U$. So, if
$U=\left\{  u_{1},...,u_{N}\right\}  $ then $\left\vert U\right\vert =N$.

\item Given sets $U,W$ we define the \emph{set difference} $U\backslash W$ by
\[
U\backslash W=\left\{  u:u\in U\text{ and }u\not \in W\right\}  .
\]

\item Given a reference set $U=\left\{  u_{1},...,u_{N}\right\}  $, any set
$W\subseteq U$ can be described by its \emph{indicator vector} $\mathbf{w}%
=\left[  w_{1},w_{2},...,w_{N}\right]  $ where%
\[
\forall n\in\left\{  1,...,N\right\}  :w_{n}=\left\{
\begin{array}
[c]{ll}%
1 & \text{iff }u_{n}\in W,\\
0 & \text{iff }u_{n}\not \in W.
\end{array}
\right.
\]

\end{enumerate}

\noindent We use the following graph theoretic definitions and notation.

\begin{enumerate}
\item A \emph{graph} is a pair $G=\left(  V,E\right)  $, where the
\emph{vertex set} is $V=\left\{  x_{1},x_{2},...,x_{N}\right\}  $ and the
\emph{edge set }$\ $is%
\[
E\subseteq\left\{  \left\{  x,y\right\}  :x,y\in V\text{, }x\neq y\right\}  ;
\]

a \emph{directed graph} is a pair $G=\left(  V,E\right)  $, where the
\emph{vertex set} is $V=\left\{  1,2,...,N\right\}  $ and the \emph{edge set
}$\ $is%
\[
E\subseteq\left\{  \left(  x,y\right)  :x,y\in V\text{, }x\neq y\right\}  .
\]

In both of the above cases we usually (but not always)\ take the node set to
be $V=\left\{  1,2,...,N\right\}  $.

\item For graph $G=\left(  V,E\right)  $ and vertex $x\in V$:%
\begin{align*}
\text{the \emph{neighborhood} of }x\text{ is}  &  :N\left(  x\right)
=\left\{  y:\left\{  x,y\right\}  \in E\right\}  ,\\
\text{the \emph{closed neighborhood} of }x\text{ is}  &  :N\left[  x\right]
=N\left(  x\right)  \cup\left\{  x\right\}  .
\end{align*}

\item For \emph{graph} $G=\left(  V,E\right)  $ with $\left\vert V\right\vert
=N$, the \emph{adjacency matrix} of $G$ is an $N\times N$ matrix $A$ defined
by%
\[
\forall x,y\in V:A_{xy}=\left\{
\begin{array}
[c]{ll}%
1 & \text{iff }\left\{  x,y\right\}  \in E,\\
0 & \text{iff }\left\{  x,y\right\}  \not \in E.
\end{array}
\right.
\]
For \emph{directed graph} $G=\left(  V,E\right)  $ with $\left\vert
V\right\vert =N$, the \emph{adjacency matrix} of $G$ is an $N\times N$ matrix
$A$ defined by%
\[
\forall x,y\in V:A_{xy}=\left\{
\begin{array}
[c]{ll}%
1 & \text{iff }\left(  x,y\right)  \in E,\\
0 & \text{iff }\left(  x,y\right)  \not \in E.
\end{array}
\right.
\]

A graph $G=\left(  V,E\right)  $ is uniqely specified by its adjacency matrix;
hence in what follows we will sometimes say \textquotedblleft the graph
$A$\textquotedblright. The same is true of a directed graph.

\item For \emph{graph} $G=\left(  V,E\right)  $ with $\left\vert V\right\vert
=N$, a \emph{visibility matrix} for $G$ is an $N\times N$ matrix $B$ defined
by%
\[
\forall x,y\in V:B_{xy}=\left\{
\begin{array}
[c]{ll}%
1 & \text{iff }y\text{ can be seen from }x,\\
0 & \text{iff }y\text{ cannot be seen from }x.
\end{array}
\right.
\]

\end{enumerate}

\section{The LVGS\ Problem and Algorithm\label{sec03}}

We start with an informal description of the LVGS\ problem.

\begin{enumerate}
\item We are given: the graph $G=\left(  V,E\right)  $, specified by its
adjacency matrix $A$, and a visibility matrix $B$ for $G$.

\item At time $t=0$ we place each of $K$ searchers at a graph vertex (a vertex
can contain any number of searchers).\ It is also assumed that a target is
located at some unknown graph vertex.

\item At discrete time steps $t\in\left\{  1,2,\ldots\right\}  $ we can move
each searcher to a vertex neighboring his current position.

\item The following assumptions are made for each time step $t\in\left\{
0,1,2,\ldots\right\}  .$

\begin{enumerate}
\item The visibility matrix $B$ is similar to the adjacency matrix $A$. Just
like an agent located at vertex $x$ can \emph{move} to vertex $y$ iff
$A_{xy}=1$, similarly, an agent located at vertex $x$, can \emph{see} vertex
$y$ (and its contents) iff $B_{xy}=1$.

\item The target will always move to a vertex $z$ which is not seen by any
searcher, \emph{provided there is a path from his current vertex to }$z$
\emph{which does not pass through a  searcher-visible vertex}.

\item The process ends when the target is in a vertex which can be observed by
the searchers (the target is observed, the graph is \textquotedblleft
cleared\textquotedblright).
\end{enumerate}

\item Our goal is to clear the graph in the shortest possible number of steps.
\end{enumerate}

\noindent To provide a precise mathematical formulation of the above
description, we introduce the following.

\begin{definition}
\label{prp0302}\normalfont For each $t\in\left\{  1,2,\ldots\right\}  $ we
define the \emph{position vector} by%
\[
\mathbf{x}(t)=[x_{1}(t),\ldots,x_{K}(t)],
\]
where $x_{k}(t)\in V$ is the position of the $k$-th searcher at time $t$. For
all $t\geq1$, we have $\mathbf{x}(t)\in\mathbf{X}=V^{K}$.
\end{definition}

\begin{definition}
\normalfont We also introduce the \emph{null position vector }$\lambda$; hence
we write $\mathbf{x}(0)=\lambda$ to indicate that, at $t=0$, no searchers have
been placed on the graph.
\end{definition}

\begin{definition}
\label{prp0303}\normalfont If a vertex \emph{may} contain the target, the
vertex is called \emph{dirty}, otherwise it is called \emph{clear}.
\end{definition}

When a vertex $x$ is seen by a searcher it is \emph{cleared}; it remains clear
as long as there is no \textquotedblleft free\textquotedblright\ (i.e.,
invisible to the searchers) path from $x$ to a dirty vertex. If, at some time,
a path from a clear vertex $x$ to a dirty vertex $y$ becomes invisible to the
searchers, then $x$ becomes dirty again (is \emph{recontaminated}). We will
denote the set of all dirty (resp. clear) vertices at time $t$ by $D\left(
t\right)  $ (resp. by $C\left(  t\right)  $). We can describe sets by their
\emph{indicator vectors}; we will denote the indicator vector of $D\left(
t\right)  $ by
\[
\mathbf{d}\left(  t\right)  =[d_{1}\left(  t\right)  ,d_{2}\left(  t\right)
,...,d_{N}\left(  t\right)  ].
\]
More precisely, we have the following.

\begin{definition}
\label{prp0304}\normalfont For each $t\in\left\{  0,1,2,\ldots\right\}  $ we
define the \emph{contamination vector} by
\[
\mathbf{d}\left(  t\right)  =[d_{1}\left(  t\right)  ,d_{2}\left(  t\right)
,...,d_{N}\left(  t\right)  ]
\]
where%
\[
\forall t,n:d_{n}\left(  t\right)  =\left\{
\begin{array}
[c]{ll}%
1 & \text{iff vertex }n\text{ is dirty at time }t\text{,}\\
0 & \text{iff vertex }n\text{ is clear at time }t\text{.}%
\end{array}
\right.
\]
We set $\mathbf{d}\left(  0\right)  =\left[  1,...,1\right]  $, i.e., at $t=0$
all vertices are dirty.
\end{definition}

\noindent Next we define the \emph{state vector}, which contains all the
information relevant to the graph search.

\begin{definition}
\label{prp0305}\normalfont For each $t\in\left\{  0,1,2,\ldots\right\}  $ we
define the \emph{state vector} by
\[
\mathbf{z}(t)=\left[  x_{1}(t),...,x_{K}(t),d_{1}(t),...,d_{N}(t)\right]
=\left[  \mathbf{x,d}\right]  .
\]
For all $t\geq1$, we have $\mathbf{x}(t)\in V^{K}\times\left\{  0,1\right\}
^{N}$. Taking in account $t=0$, the \emph{state space} (set of all possible
states) is
\[
\mathbf{Z}=\left(  V^{K}\times\left\{  0,1\right\}  ^{N}\right)  \cup\left\{
\left[  \lambda,1,...,1\right]  \right\}  .
\]

\end{definition}

It is assumed that, at every time $t$, the \textquotedblleft search
planner\textquotedblright\ knows $\mathbf{z}(t)$; i.e., he knows the location
of all searchers and the \emph{possible} locations of the target.

\begin{definition}
\label{prp0306}\normalfont The \emph{clear states} are the elements of
$\mathbf{Z}$ which have the form
\[
\mathbf{z}=\left[  \mathbf{x,}\left[  0,...,0\right]  \right]  ,
\]
and the set of all clear states will be denoted by $\mathbf{Z}_{C}$. The set
of all dirty \emph{states is }%
\[
\mathbf{Z}_{D}=\mathbf{Z}\backslash\mathbf{Z}_{C}%
\]
and the \emph{dirty states }are the elements of $\mathbf{Z}_{D}$.
\end{definition}

\begin{definition}
\label{prp0307}\normalfont The \emph{control vector} is
\[
\mathbf{u}(t)=[u_{1}(t),\ldots,u_{K}(t)]
\]
where $u_{k}(t)$ denotes the position to which the $k$-th searcher will be
moved at time $t+1$ (so at $t+1$ the $k$-th searcher will be located at
$x_{k}(t+1)=u_{k}(t)$). Of course moves must be legal, i.e.,
\begin{equation}
\forall t,k:\left\{  x_{k}(t),u_{k}(t)\right\}  \in{E}. \label{eq0301}%
\end{equation}

\end{definition}

To specify the \textquotedblleft\emph{search evolution equation}%
\textquotedblright\ we first define \emph{star-multiplication} $\ast$ as follows.

\begin{definition}
\label{prp0309}\normalfont Given the $L\times M$ matrix $P$ and the $M\times
N$ matrix $Q$ we define the $L\times N$ matrix $P\ast Q$ as follows%
\[
\forall l\in\{1,2,\ldots,L\},n\in\{1,2,\ldots,N\}:\left(  P\ast Q\right)
_{ln}=\max_{m=1,2,...M}[\min(P_{lm},Q_{mn})]\text{.}%
\]

\end{definition}

When the target's speed is equal to $s$ (i.e., at every time step he can
traverse $s$ edges)\ and assuming that no searchers exist in the graph, we
have the following equation for the evolution of the dirty
set\footnote{\noindent If the target is arbitrarily fast, then it suffices to
set $s=N$, the number of vertices in the graph.}
\[
\mathbf{d}(t+1)=\mathbf{d}(t)\ast A\underset{s\text{ times}}{\ast\cdots\ast}A
\]
Now suppose a single searcher is placed at vertex $u_{1}$ . The searcher
blocks certain paths which, if taken by the target, would result in his being
observed. In particular, the target will not move into any vertex which
belongs to the visibility field of the searcher; similarly, the target will
not move out of any vertex in the visibility field. Similar rules hold in case
$K$ searchers are located in the graph.

To model this situation, we introduce the \emph{modified adjacency matrix}
$\overline{A}\left(  \mathbf{u}\right)  $. It describes the vertices into
which the target can move without being observed by searchers located at
positions $\mathbf{u}=\left[  u_{1},...,u_{K}\right]  $. In other words,
$\overline{A}\left(  \mathbf{u}\right)  $ is the adjacency matrix of the graph
$\overline{G}\left(  \mathbf{u}\right)  $ obtained by removing from the graph
$G$ the vertices which lie in the visibility region of the searchers (and all
their incident edges).

\begin{definition}
\label{prp0310}\normalfont The \emph{modified adjacency matrix} $\overline
{A}\left(  \mathbf{u}\right)  $, where $\mathbf{u=}\left[  u_{1}%
,...,u_{K}\right]  $, is defined as follows%
\[
\forall x,y:\overline{A}_{xy}\left(  \mathbf{u}\right)  =\left\{
\begin{array}
[c]{ll}%
A_{xy}\text{ } & \text{iff: }\forall k\in\left\{  1,2,...,K\right\}  :\left\{
x,y\right\}  \cap N\left[  u_{k}\right]  =\emptyset\text{,}\\
0 & \text{otherwise.}%
\end{array}
\right.
\]

\end{definition}

Putting all of the above pieces together we can write an equation which
specifies the next state of the graph search, given the current state and control.

\begin{definition}
\label{prp0308}\normalfont The \emph{graph search evolution equation} (or
simply the \emph{evolution equation}) is%
\begin{align}
\mathbf{x}(t+1)  &  =\mathbf{u}(t),\label{eq0302}\\
\mathbf{d}(t+1)  &  =\mathbf{d}(t)\ast\overline{A}(\mathbf{u}(t))\underset
{s\text{ times}}{\ast\cdots\ast}\overline{A}(\mathbf{u}(t)). \label{eq0303}%
\end{align}

\end{definition}

\bigskip

\bigskip

\noindent Now we are ready to formally define a \emph{preliminary version} of
the LVGS\ problem as follows.

\begin{tcolorbox}[breakable,colback=white,colframe=black,width=\dimexpr \textwidth \relax]
\begin{problem}
\normalfont \textbf{(Preliminary LVGS\ Problem) }Given $A$ and $B$ choose,
subject to (\ref{eq0302})-(\ref{eq0303}), the minimum $T$ and respective
$\mathbf{u}(0),...,\mathbf{u}(T-1)$ such that $\left[  \mathbf{x}(T),\mathbf{d}(T)\right]
\in \mathbf{Z}_{C}$.
\end{problem}
\end{tcolorbox}

This problem can be solved by graph theoretic methods. Consider the directed
graph $G^{\prime}=\left(  V^{\prime},E^{\prime}\right)  $ where the vertices
are the previously defined states (i.e., $V^{\prime}=\mathbf{Z}$)\ and the
edges correspond to valid state (i.e., an edge $\left(  \mathbf{z}%
_{1},\mathbf{z}_{2}\right)  $ denotes that we can make a transition from
$\mathbf{z}_{1}=\left(  \mathbf{x}_{1},\mathbf{d}_{1}\right)  $ to
$\mathbf{z}_{2}=\left(  \mathbf{x}_{2},\mathbf{d}_{2}\right)  $).\ Then we can
solve the LVGS\ problem by finding a shortest path from the vertex $\left(
\lambda,\left[  1,...,1\right]  \right)  $\ to some vertex of the form
$\left(  \mathbf{x},\left[  0,...,0\right]  \right)  \in\mathbf{Z}_{C}$, which
can be solved by, e.g., Dijkstra's algorithm.

However, in the above formulation the size of $\mathbf{Z}$ is $\left\vert
\mathbf{Z}\right\vert =\left\vert V\right\vert \cdot2^{\left\vert V\right\vert
}+1$ which grows exponentially. For instance, when the original $G$ has
$\left\vert V\right\vert =10$ vertices and we employ a single searcher
($K=1$), then $\left\vert \mathbf{Z}\right\vert =10\cdot2^{10}+1=\allowbreak
10\,241$; with $\left\vert V\right\vert =20$ vertices and a single searcher we
have $\left\vert \mathbf{Z}\right\vert =20\cdot2^{20}+1=\allowbreak
20\,971\,521$.

Consequently, in what follows we will introduce a different approach (inspired
from \cite{Guibas1997,Guibas1999}) which can achieve the same goal (reaching
an all-clear state)\ using a much smaller directed graph $G_{I}$, the
\textquotedblleft\emph{information state graph}\textquotedblright; $G_{I}$
describes how the contamination of $G$ (i.e., the possible locations of the
target) changes as the searcher moves in $G$.

To describe the structure of $G_{I}$, we consider the evolution of
contamination as the searchers move through $G$. For the sake of simplicity we
will consider the case of a single searcher (the extension to the case of many
searchers is straightforward). Supposing that the searcher is located on some
vertex $x$, then

\begin{enumerate}
\item we have a \emph{visible} component, defined to be the subgraph induced
by the vertices which are visible to the searcher;

\item unless the search problem is trivial, we will also have some
\emph{invisible} components, defined to be the connected components of the
graph obtained by removing the visible component; by this definition, the
vertices of the invisible components cannot be seen by the searcher.
\end{enumerate}

\noindent Hence, given $A$ and $B$, and for every $x\in V$, we have a
\emph{visible component} and one or more \emph{invisible components}%
\footnote{Strictly speaking, the components are subgraphs of $G$ but, since
$G$ is given, they are fully specified by their vertex sets. Hence
\textquotedblleft components\textquotedblright\ will also be used to refer to
the corresponding vertex sets.}. By definition, the vertices in the searcher's
visible component are clear. On the other hand, some of the vertices of the
invisible components may be contaminated. Suppose that at some time the
searcher is located at vertex $x$, resulting at the partition of $V$ into $M$
invisible components (we can omit the visible component, since it will always
be clear); the contamination of the components is encoded by a vector
$\widehat{\mathbf{d}}=\left[  \widehat{d}_{1},...,\widehat{d}_{M}\right]  $,
where $\widehat{d}_{m}$ equals $0$ (resp. $1$) iff the $m$-th component is
clear (resp. dirty).\footnote{Note that the size of $\widehat{\mathbf{d}}$ is
not fixed since, for different searcher positions, we will have a different
number of visibility components.} The pair $\widehat{\mathbf{z}}%
=(x,\widehat{\mathbf{d}})$ will be called the \emph{information state }and
gives us all the needed information about the current state of the search.

\begin{example}
\label{prop0401}\normalfont Let us clarify the above ideas with an example.
Consider the tree of Figure \ref{fig0401}. Assume a single searcher with a
visibility range of $L=1$ (i.e., he can see all the vertices which are at
distance at most one from his current position). Consequently, the visibility
matrix is%
\[
B=\left[
\begin{array}
[c]{ccccccc}%
1 & 1 & 1 & 0 & 0 & 0 & 0\\
1 & 1 & 0 & 1 & 1 & 0 & 0\\
1 & 0 & 1 & 0 & 0 & 1 & 1\\
0 & 1 & 0 & 1 & 0 & 0 & 0\\
0 & 1 & 0 & 0 & 1 & 0 & 0\\
0 & 0 & 1 & 0 & 0 & 1 & 0\\
0 & 0 & 1 & 0 & 0 & 0 & 1
\end{array}
\right]
\]
Assume the searcher is located on vertex $1$; then he can see vertices $1,2$
and $3$. Hence, as can be seen in Figure \ref{fig0402}, the visible component
is $V_{1}=\left\{  1,2,3\right\}  $ and the invisible components are
$V_{2}=\left\{  4\right\}  $,$\ V_{3}=\left\{  5\right\}  $, $V_{4}=\left\{
6\right\}  $, and $V_{5}=\left\{  7\right\}  $.
\end{example}

\noindent\begin{minipage}[c]{0.4 \textwidth}
\medskip
\centering
\includegraphics[scale=0.5]{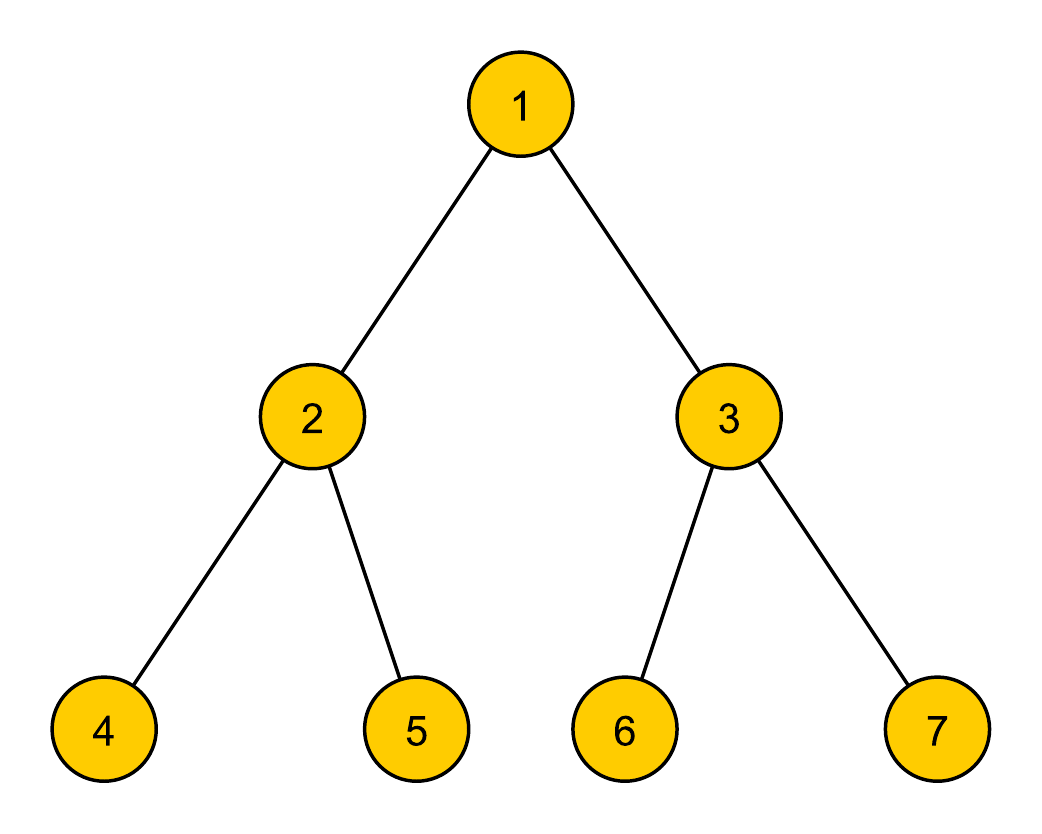}
\captionof{figure}{A graph  to be searched.}
\label{fig0401}
\medskip
\end{minipage}\qquad\begin{minipage}[c]{0.4 \textwidth}
\medskip
\centering
\includegraphics[scale=0.5]{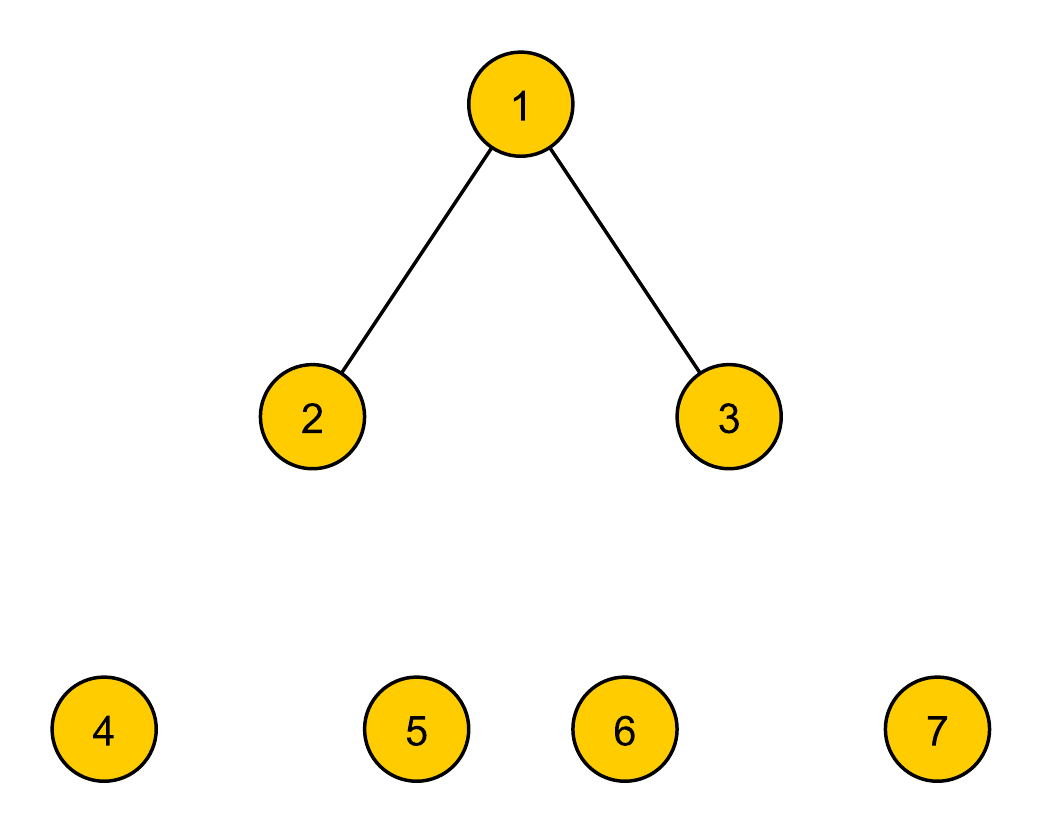}
\captionof{figure}{Visible and invisible components of the graph when the searcher is at vertex 4
and has straight line visibility (he  can see vertices 1, 2 and 3).}
\label{fig0402}
\medskip
\end{minipage}

\noindent Now suppose the searcher performs the following search schedule:
$\mathbf{u}(0)=1\rightarrow\mathbf{u}(1)=2\rightarrow\mathbf{u}%
(2)=1\rightarrow$ $\mathbf{u}(3)=3$. Then we have the following evolution of
clear and contaminated components (in the third column, the visible component
verices are listed in italics).
\begin{align*}
&
\begin{tabular}
[c]{|l|l|l|l|l|}\hline
Time & Pursuer Location & Visibility Components & Contamination Vector &
Information state\\\hline
$0$ & $\lambda$ & $\left\{  1,2,...,7\right\}  $ & $\left(  1\right)  $ &
$\left(  \lambda,\left(  1\right)  \right)  $\\\hline
$1$ & $1$ & $\left\{  \emph{1,2,3}\right\}  ,\left\{  4\right\}  ,\left\{
5\right\}  ,\left\{  6\right\}  ,\left\{  7\right\}  $ & $\left(
1,1,1,1\right)  $ & $\left(  1,\left(  1,1,1,1\right)  \right)  $\\\hline
$2$ & $2$ & $\left\{  \emph{1,2,4,5}\right\}  ,\left\{  3,6,7\right\}  $ &
$\left(  1\right)  $ & $\left(  2,\left(  1\right)  \right)  $\\\hline
$3$ & $1$ & $\left\{  \emph{1,2,3}\right\}  ,\left\{  4\right\}  ,\left\{
5\right\}  ,\left\{  6\right\}  ,\left\{  7\right\}  $ & $\left(
0,0,1,1\right)  $ & $\left(  1,\left(  0,0,1,1\right)  \right)  $\\\hline
$4$ & $3$ & $\left\{  \emph{1,3,6,7}\right\}  ,\left\{  2,4,5\right\}  $ &
$\left(  0\right)  $ & $\left(  3,\left(  0\right)  \right)  $\\\hline
\end{tabular}
\ \ \ \ \ \ \\
&  \text{Table 1. Visibility components, contamination vectors and information
states for a walk through }\\
&  \text{the binary tree of depth 2, with visibility range }L=1\text{.}%
\end{align*}
The entries in the above table can be explained as follows.

\begin{enumerate}
\item At $t=0$ the searcher is outside the graph. The single invisible
component contains all graph vertices.

\item At $t=1$ the searcher moves to vertex $1$ and sees vertices $2,3$. All
of these form a single clear component; each of vertices $4,5,6,7$ forms a
separate component which is contaminated (may contain the target); hence the
contamination vector is $\left(  1,1,1,1\right)  $.

\item At $t=2$ the searcher moves to vertex $2$. Now he sees vertices
$1,2,4,5$, all of which form a single clear component and the vertices $3,6,7$
forms a single contaminated component. Note that vertices $4$ and $5$ have
been cleared and vertex $3$ has been recontaminated. Hence the contamination
vector is $\left(  1\right)  $.

\item At $t=3$ the searcher moves back to vertex $1$. The components are the
same as at time $t=1$, but now they are in a different contamination state,
namely, $\left(  0,0,1,1\right)  $. This is so because vertices $4$ and $5$
were previously clear and, though no longer directly visible, cannot be
recontaminated because the target cannot move from $6$ or $7$ to $4$ or $5$
without passing through the currently visible component $\left\{
1,2,3\right\}  $.

\item Finally, at $t=4$ the searcher moves to vertex $3$. The components now
are $\left\{  1,3,6,7\right\}  $ (which is visible) and $\left\{
2,4,5\right\}  $ (which was clear and cannot be recontaminated). Hence the
contamination vector is $\left(  0\right)  $ and the graph has been cleared.
\end{enumerate}

\noindent Note how the number of components and the vertices they include
change during the search; also note that the same components can, at different
times, be clear or contaminated depending on the search history.

\bigskip

\bigskip

\begin{example}
\label{prop0401a}\normalfont The previous example employed \textquotedblleft
distance-based visibility\textquotedblright. Now we present an example with
\textquotedblleft straight-line visibility\textquotedblright. Consider the
graph of Figure \ref{fig0401a} in which each vertex is supposed to be actually
located in the position indicated in the figure. Assume a single searcher with
straight-line visibility (i.e., he can see all the vertices which lie in a
straight-line path passing from his current vertex). Consequently, the
visibility matrix is
\[
B=\left[
\begin{array}
[c]{cccccccccccc}%
1 & 1 & 0 & 0 & 0 & 0 & 0 & 0 & 0 & 0 & 0 & 0\\
1 & 1 & 1 & 0 & 0 & 0 & 0 & 0 & 0 & 0 & 0 & 0\\
0 & 1 & 1 & 1 & 0 & 0 & 0 & 0 & 1 & 0 & 0 & 0\\
0 & 0 & 1 & 1 & 1 & 0 & 1 & 1 & 1 & 0 & 0 & 0\\
0 & 0 & 0 & 1 & 1 & 1 & 1 & 1 & 0 & 0 & 0 & 0\\
0 & 0 & 0 & 0 & 1 & 1 & 0 & 0 & 0 & 0 & 0 & 0\\
0 & 0 & 0 & 1 & 1 & 0 & 1 & 1 & 0 & 0 & 0 & 0\\
0 & 0 & 0 & 0 & 1 & 0 & 1 & 1 & 0 & 0 & 0 & 0\\
0 & 0 & 1 & 1 & 0 & 0 & 0 & 0 & 1 & 1 & 1 & 1\\
0 & 0 & 0 & 0 & 0 & 0 & 0 & 0 & 1 & 1 & 1 & 1\\
0 & 0 & 0 & 0 & 0 & 0 & 0 & 0 & 1 & 1 & 1 & 1\\
0 & 0 & 0 & 0 & 0 & 0 & 0 & 0 & 1 & 1 & 1 & 1
\end{array}
\right]
\]
Assume the searcher is at vertex 4; then he can see vertices 3, 4, 5, 7, 8 and
9. Hence, as can be seen in Figure \ref{fig0402a}, the visible component is
$V_{1}=\left\{  3,4,5,7,8,9\right\}  $ and the invisible components are
$V_{2}=\left\{  1,2\right\}  $,$\ V_{3}=\left\{  6\right\}  $, $V_{4}=\left\{
10\right\}  $, and $V_{5}=\left\{  11,12\right\}  $.
\end{example}

\noindent\begin{minipage}[c]{0.4 \textwidth}
\medskip
\centering
\includegraphics[scale=0.5]{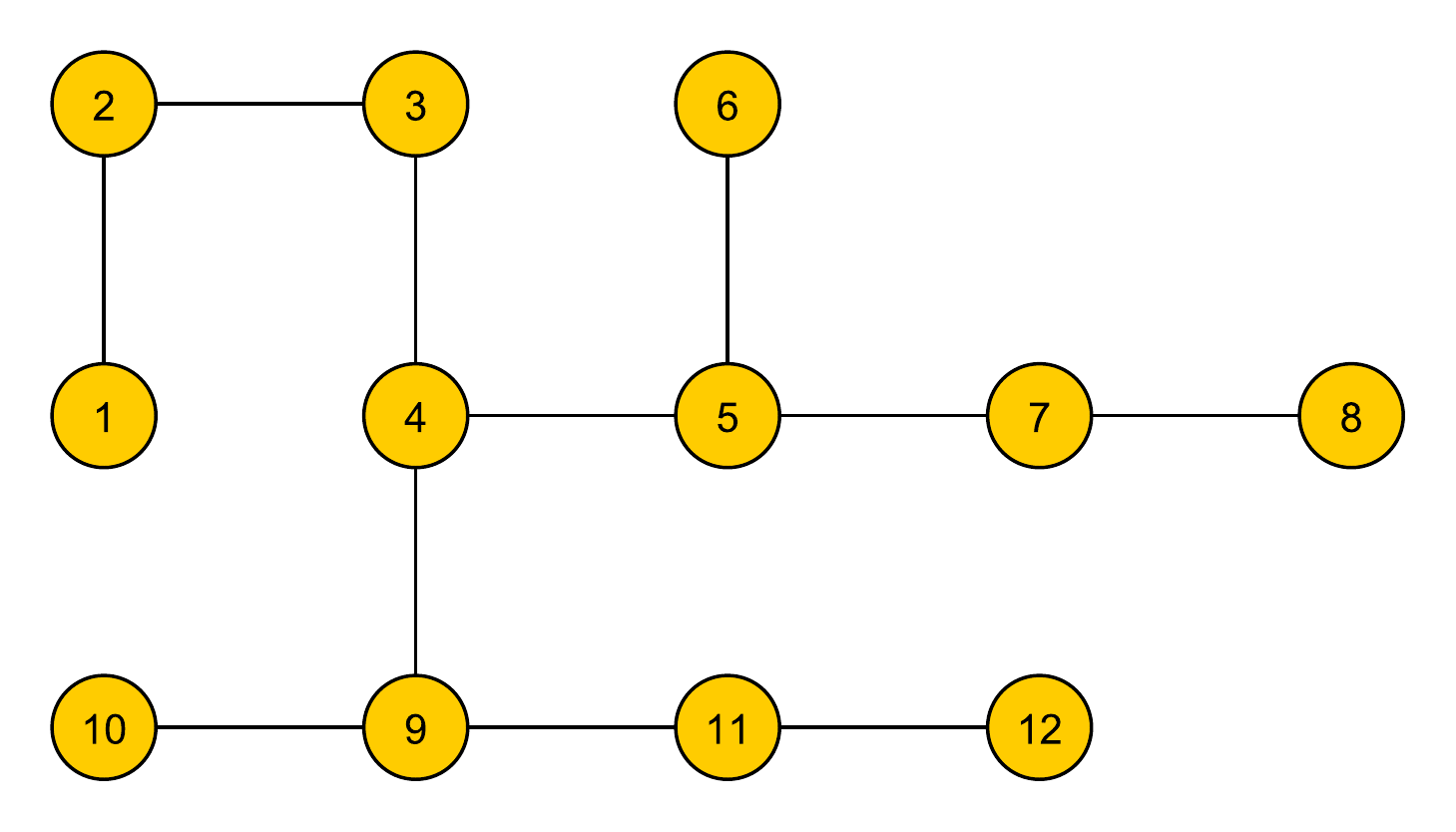}
\captionof{figure}{A graph to be searched.}
\label{fig0401a}
\medskip
\end{minipage}\qquad\begin{minipage}[c]{0.4 \textwidth}
\medskip
\centering
\includegraphics[scale=0.5]{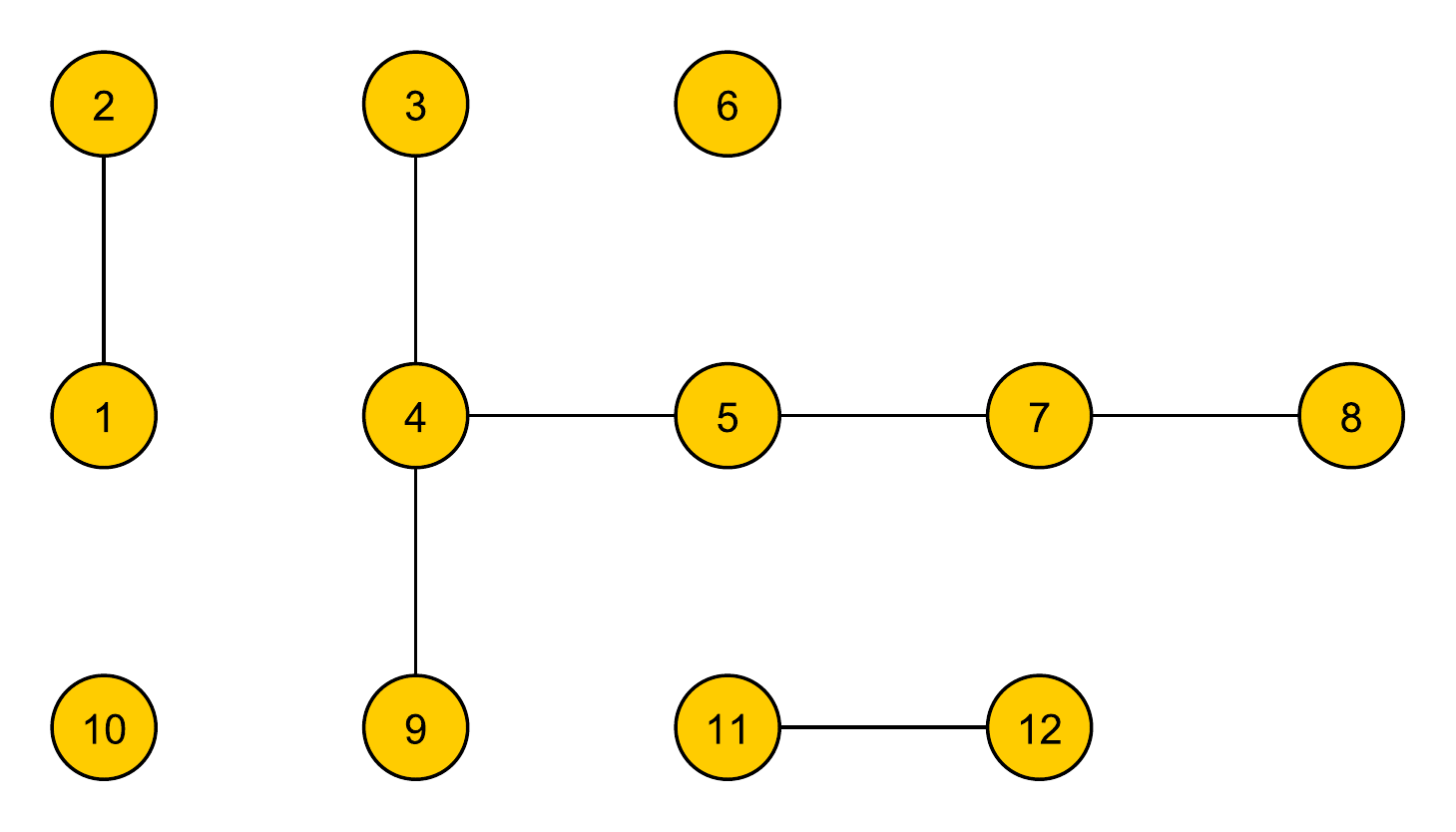}
\captionof{figure}{Visible and invisible components of the graph when the searcher is at vertex 4
(he  can see vertices 3, 4, 5, 7, 8 and 9).}
\label{fig0402b}
\medskip
\end{minipage}

\noindent Now suppose the searcher performs the following search schedule:
$\mathbf{u}(0)=1\rightarrow\mathbf{u}(1)=2\rightarrow\mathbf{u}%
(2)=3\rightarrow$ $\mathbf{u}(3)=4$, $\mathbf{u}(4)=9\rightarrow
\mathbf{u}(5)=4\rightarrow\mathbf{u}(6)=5$. \ Then we have the following
evolution of clear and contaminated components (in the third column, the
visible component verices are listed in italics).
\begin{align*}
&
\begin{tabular}
[c]{|l|l|l|l|l|}\hline
Time & Pursuer Location & Visibility Components & Contamination Vector &
Information state\\\hline
$0$ & $\lambda$ & $\left\{  1,2,...,12\right\}  $ & $\left(  1\right)  $ &
$\left(  \lambda,\left(  1\right)  \right)  $\\\hline
$1$ & $1$ & $\left\{  \emph{1,2}\right\}  ,\left\{  3,4,...,12\right\}  $ &
$\left(  1\right)  $ & $\left(  1,\left(  1\right)  \right)  $\\\hline
$2$ & $2$ & $\left\{  \emph{1,2,3}\right\}  ,\left\{  4,5,...,12\right\}  $ &
$\left(  1\right)  $ & $\left(  2,\left(  1\right)  \right)  $\\\hline
$3$ & $3$ & $\left\{  1\right\}  ,\left\{  \emph{2,3,4,9}\right\}  ,\left\{
5,6,7,8\right\}  ,\left\{  10\right\}  ,\left\{  11,2\right\}  $ & $\left(
0,1,1,1\right)  $ & $\left(  3,\left(  0,1,1,1\right)  \right)  $\\\hline
$4$ & $4$ & $\left\{  1,2\right\}  ,\left\{  \emph{3,4,5,7,8,9}\right\}
,\left\{  6\right\}  ,\left\{  10\right\}  ,\left\{  11,12\right\}  $ &
$\left(  0,1,1,1\right)  $ & $\left(  4,\left(  0,1,1,1\right)  \right)
$\\\hline
$5$ & $9$ & $\left\{  1,2\right\}  ,\left\{  \emph{3,4,9,10,11,12}\right\}
,\left\{  5,6,7,8\right\}  $ & $\left(  0,1\right)  $ & $\left(  9,\left(
0,1\right)  \right)  $\\\hline
$6$ & $4$ & $\left\{  1,2\right\}  ,\left\{  \emph{3,4,5,7,8,9}\right\}
,\left\{  6\right\}  ,\left\{  10\right\}  ,\left\{  11,12\right\}  $ &
$\left(  0,1,0,0\right)  $ & $\left(  4,\left(  0,1,0,0\right)  \right)
$\\\hline
$7$ & $5$ & $\left\{  1,2,3\right\}  ,\left\{  \emph{4,5,6,7,8}\right\}
,\left\{  9,10,11,12\right\}  $ & $\left(  0,0\right)  $ & $\left(  5,\left(
0,0\right)  \right)  $\\\hline
\end{tabular}
\ \ \ \ \ \ \\
&  \text{Table 2. Visibility components, contamination vectors and information
states for a walk through }\\
&  \text{the graph with straight line visibility.}%
\end{align*}
The reader can check that the search schedule clears the graph, which is also
seen by the final information state $\left(  5,\left(  0,0\right)  \right)  $.

\bigskip

\bigskip

We denote by $\widehat{\mathbf{Z}}$ the set of all information states
$(x,\widehat{\mathbf{d}})$ and by $\widehat{\mathbf{Z}}_{C}$ the set of all
clear information states $(x,\left[  0,...,0\right]  )$. There exists a
\emph{state transition function} $F:\widehat{\mathbf{Z}}_{C}\times
V\rightarrow\widehat{\mathbf{Z}}_{C}$ which maps the current information state
and searcher move to the next information state:%
\begin{equation}
\widehat{\mathbf{z}}\left(  t+1\right)  =F\left(  \widehat{\mathbf{z}}\left(
t\right)  ,u\left(  t\right)  \right)  . \label{eq0304}%
\end{equation}
While it is difficult to give a general description for $\widehat{\mathbf{Z}}$
and $F$, \ they can be easily constructed algorithmically for a given graph
$G$. The information graph $G_{I}=\left(  V_{I},E_{I}\right)  $, has
$V_{I}=\widehat{\mathbf{Z}}$ and $E_{I}$ consists of the state transitions
which are described by the state transition function $F$.

\begin{example}
\label{prop0401b} \normalfont In Figure \ref{figPath5GI}.a we present the
information graph corresponding to the graph of Example \ref{prop0401}, with
visibility range $L=1$. The information graph has $\left\vert V_{I}\right\vert
=36$\textbf{ }vertices; compare this to the \ previously mentioned $V^{\prime
}$\ which has $\left\vert V^{\prime}\right\vert =\left\vert V\right\vert
\cdot2^{\left\vert V\right\vert }+1=7\cdot2^{7}+1=\allowbreak897$ vertices.
\ \noindent In this case, since the information graph is of low complexity, it
is easy to check by inspection that the shortest clearing presented in Table
1\ also appears in the information graph.

In Figure \ref{figPath5GI}.b we see the information graph corresponding to
Example \ref{prop0401a}. The information graph has $\left\vert V_{I}%
\right\vert =76$\textbf{ }vertices; compare this to the \ previously mentioned
$V^{\prime}$\ which has $\left\vert V^{\prime}\right\vert =\left\vert
V\right\vert \cdot2^{\left\vert V\right\vert }=12\cdot2^{12}+1=\allowbreak
49\,153$ vertices.

\noindent\begin{minipage}[c]{\textwidth}
\medskip
\noindent
\includegraphics[scale=0.60]{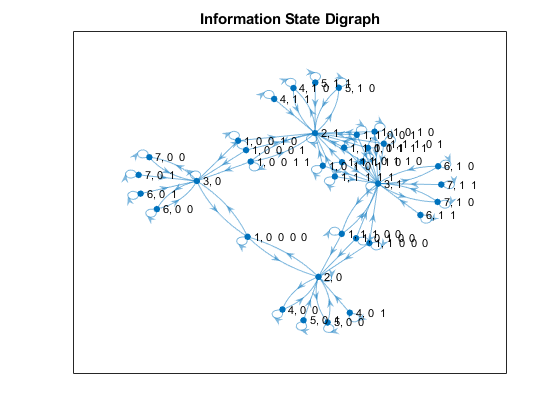}\includegraphics[scale=0.60]{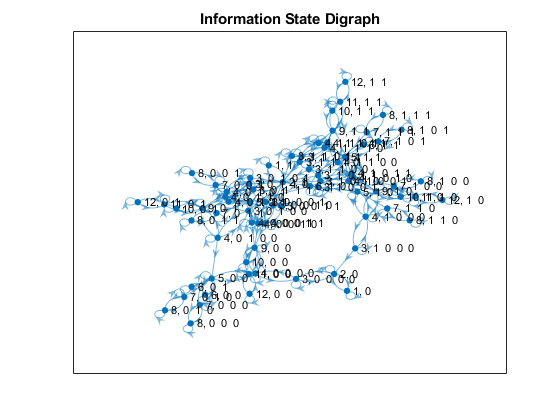}
\captionof{figure}{(a) The information digraph for Example \ref{prop0401},  with visibility range $L=1$;
(b) The information digraph for Example \ref{prop0401a},  with straight-line visibility.}
\label{figPath5GI}
\medskip
\end{minipage}

\end{example}

\bigskip

\noindent Now we can formulate the LVGS problem as follows.

\begin{tcolorbox}[breakable,colback=white,colframe=black,width=\dimexpr \textwidth \relax]
\begin{problem}
\normalfont \textbf{(Main LVGS\ Problem) }Given $A$ and $B$ choose, subject to
(\ref{eq0304}), the minimum $T$ and respective $u(0),...,u(T-1)$ such that
$\left[  x(T),\widehat{\mathbf{d}}(T)\right]  \in \widehat{\mathbf{Z}}_{C}$.
\end{problem}
\end{tcolorbox}

The problem reduces to finding, in the graph $G_{I}$, a shortest path from the
unique all-dirty state $\left(  \lambda,\left[  1\right]  \right)  $ (i.e.,
when the searcher has not yet been placed in the graph and there is a single
invisible and dirty component) to some all-clear state $\left(  \mathbf{x,}%
\left[  0,...,0\right]  \right)  \in\mathbf{Z}_{C}$. Note that we also have to
\emph{construct} the information graph $G_{I}$ and the transition function
$F$. The following LVGS algorithm constructs the graph and finds a shortest
path. \noindent The pseudocode is rather self-explanatory; here are a few clarifications.

\begin{enumerate}
\item Lines 3-7: for each vertex in $G$, a group of vertices is added to
$G_{I}$, each corresponding to a different possible information state.

\item Lines 8-15: edges are added to $G_{I}$ by following the searcher's
possible transitions between adjacent $G_{I}$ vertices, by the routine
\textbf{newInfoState}.\footnote{Actually we have two versions of this
routine:\ (a)\ the first version corresponds to sequential moves (i.e., first
the searcher moves with unit speed and the the target moves with unbounded
speed); (b)\ the second version implements concurrent moves by both searcher
and target and is somewhat more complicated (details are omitted, in the
interest of simplicity).}

\item Lines 6-20: using Dijkstra's algorithm, we find a shortest path between
the all-dirty vertex and every all-clear vertex of $G_{I}$.\footnote{For very
large $G_{I}$ we can instead use a depth-first search starting at $\left(
\lambda,\left(  1\right)  \right)  $; this will find a clearing search
schedule, though not the shortest one.}

\item Lines 21-22: the algorithm returns the overall shortest path.
\end{enumerate}

This concludes our formulation and proposed solution of the LVGS problem. It
is worth emphasizing that, in both formulation and solution, the only decision
maker is the searcher; the target has been \textquotedblleft factored
out\textquotedblright\ of the problem. This is achieved by adopting a worst
case analysis, namely that every \emph{possible} contamination is actually
\emph{realized}.

\bigskip

\begin{tcolorbox}[breakable,colback=white,colframe=black,width=\dimexpr \textwidth \relax]
\begin{minipage}[c]{0.95\textwidth}
\medskip
\begin{center}
\par \textbf{The LVGS Algorithm}
\end{center}
\medskip
\begin{algorithmic}[1]
\Function{LVGS}{$A$, $B$}
\State Let $G=(V,E)$ be the graph with adjacecny matrix $A$
\For {$i\in V$}  \Comment{Build vertex set $V_I$ of $G_I=(V_I,E_I)$}
	\For {every possible $\mathbf{c}(i)$ } \Comment{$\mathbf{c}(i)$: contamination state of \(i\)}
		\State add vertex $(i,\mathbf{c}(i))$ to $V_I$
	\EndFor
\EndFor
\For {$i\in V$}  \Comment{Build edge set $E_I$ of $G_I=(V_I,E_I)$}
	\For {$j\in N[i]$}
		\For {every possible $\mathbf{c}(i)$}
			\State calculate $\mathbf{c}(j)=\mathbf{newInfoState}(A,B)$ \Comment {Using algorithm newInfoState}
			\State add edge $(i,\mathbf{c}(i))\rightarrow(j,\mathbf{c}(j))$ to $E_I$
		\EndFor
	\EndFor
\EndFor
\State Let $u=(\lambda,[1,...,1])$, the all dirty state of $V_I$
\State Let $V_I^c$ be the set of clear states of $V_I$
\For {$v\in V^c_I$}
	\State $(\text{BestCost}(v),\text{BestPath}(v))=\mathbf{Dijkstra}(G_I,u,v)$ \Comment {Using Dijkstra's algorithm}
\EndFor
\State Let $\widehat{v}=\arg \min_{v\in V^c_I}\text{BestCost}(v)$
\State \textbf{return} $(\text{BestCost}(\widehat{v}),\text{BestPath}(\widehat{v}))$
\EndFunction
\end{algorithmic}
\medskip
\end{minipage}
\end{tcolorbox}

\bigskip

\section{Experiments\label{sec04}}

In this section we give examples of the application of our algorithm to
various graph families. We only present examples for \textquotedblleft
distance-based visibility\textquotedblright, i.e., the visibility matrix
determined by the \emph{visibility range parameter} $L$. Specifically, when
the cop is located at vertex $x$, he can see the vertices $y$ which are within
$L$ edges of $x$, i.e.,
\[
B_{xy}=\left\{
\begin{array}
[c]{ll}%
1 & \text{iff }d\left(  x,y\right)  \leq L,\\
0 & \text{otherwise.}%
\end{array}
\right.
\]

\subsection{Paths\label{sec0502}}

We first apply the LVGS algorithm to paths. This serves as a \textquotedblleft
reality check\textquotedblright\ since, clearly, a single searcher can clear a
path by sweeping from one end to the other. In all our experiments the
LVGS\ algorithm produced a clearing schedule of minimum length (minimum number
of steps) which is given by the following formula: with path length $N$ and
visibility range $L$, the number of required steps is%
\[
T_{\min}=\max\left(  N-\left(  2L+1\right)  ,0\right)  .
\]

\subsection{Complete Binary Trees}

\setcounter{table}{2}

The next family of graphs we examine are complete binary trees. In Table
\ref{tabCompTrees} we see the results for trees of depth $D\in\left\{
1,...,4\right\}  $ and visibility range $L=\left\{  1,...,5\right\}  $. Listed
in every cell is the length of an optimal search trajectory and computation
time (in seconds). Note that, for $D=4$ and visibility range $L=1,$ the cell
contains $\mathbf{\infty}$, which means that the searcher cannot clear the
graph. Also note that computation times increase significantly with tree
depth, even for high visibility range. E.g., for $D=4$ and visibility range
$L=3$ the cell contains \textbf{n/a}; this means that the algorithm did not
terminate after running for 5 hours. The reason for the high execution times
is that trees of higher depth must be partitioned into many components. For
example, in the complete binary tree of depth $4$ with visibility range $L=3$
the searcher from vertex 1 can see the entire tree except for the leaves.
Since every leaf is a different component, we have 16 invisible components
which contribute to $G_{I}$ $2^{16}=65536$ vertices. This is a case in which
the computational burden of finding a shortest clearing path in $G_{I}$ is
unmanageable. \footnote{However, substituting a depth first search for
Dijktra's algorithm yields a (non-optimal) clearing path in a few seconds.}

\noindent\begin{minipage}[l]{\textwidth}
\medskip
\centering
\begin{tabular}{|c|c|c|c|c|c|}
\hline
\multirow{2}{6em}{\textbf{Tree Depth}} & \multicolumn{5}{|c|}{\textbf{Visibility Range $L$}} \\ \cline{2-6}
& \textbf{1} & \textbf{2} & \textbf{3} & \textbf{4} & \textbf{5}\\
\hline
1     & 0 & 0 & 0 & 0 & 0\\
2     & 2 & 0 & 0 & 0 & 0\\
3     & 8 & 2 & 0 & 0 & 0\\
4     & $\infty$ & 8 & n/a & 0 & 0\\
\hline
\end{tabular}
\begin{tabular}{|c|r|r|r|r|r|}
\hline
\multirow{2}{6em}{\textbf{Tree Depth}} & \multicolumn{5}{|c|}{\textbf{Visibility Range $L$}} \\ \cline{2-6}
& \textbf{1} & \textbf{2} & \textbf{3} & \textbf{4} & \textbf{5}\\
\hline
1 & 0.48  & 0.05  & 0.02  & 0.02  & 0.04  \\
2 & 3.52  & 0.15  & 0.09  & 0.01  & 0.01  \\
3 & 31.14  & 82.22  & 0.65  & 0.61  & 0.41  \\
4 & 228.65  & 3063.17  & $>$ 18300.00  & 7.13  & 8.23 \\
\hline
\end{tabular}
\captionof{table}{Complete binary trees: optimal length of clearing search schedule and
computation times in seconds.}
\label{tabCompTrees}
\medskip
\end{minipage}

\subsection{Grids}

The next type of graph we look into are grids. In Table \ref{fig0202010c} we
see search schedule lengths and computation times for $M\times N$ grids with
$M,N\in\left\{  2,...,6\right\}  $. Apparently grids are actually easier to
handle than trees.

\begin{minipage}[c]{\textwidth}
\medskip
\centering
\begin{tabular}{|c|c|c|c|c|}
\hline
\multirow{2}{5em}{\textbf{Grid Size}} & \multicolumn{4}{|c|}{\textbf{Visibility Range}} \\ \cline{2-5}
& \textbf{1} & \textbf{2} & \textbf{3} & \textbf{4} \\
\hline
2x2 & $\infty$ & 0 & 0 & 0\\
2x3 & $\infty$ & 0 & 0 & 0\\
2x4 & $\infty$ & 1 & 0 & 0\\
2x5 & $\infty$ & 2 & 0 & 0\\
2x6 & $\infty$ & 3 & 1 & 0\\
3x3 & $\infty$ & 0 & 0 & 0\\
3x4 & $\infty$ & 1 & 0 & 0\\
3x5 & $\infty$ & 2 & 0 & 0\\
3x6 & $\infty$ & 3 & 1 & 0\\
4x4 & $\infty$ & $\infty$ & 1 & 0\\
4x5 & $\infty$ & $\infty$ & 1 & 0\\
4x6 & $\infty$ & $\infty$ & 3 & 1\\
5x5 & $\infty$ & $\infty$ & 2 & 0\\
5x6 & $\infty$ & $\infty$ & 3 & 1\\
6x6 & $\infty$ & $\infty$ & $\infty$ & 3\\
\hline
\end{tabular}
\begin{tabular}{|c|r|r|r|r|}
\hline
\multirow{2}{5em}{\textbf{Grid Size}} & \multicolumn{4}{|c|}{\textbf{Visibility Range $L$}} \\ \cline{2-5}
& \textbf{1} & \textbf{2} & \textbf{3} & \textbf{4} \\
\hline
2x2 & 0.81  &   0.03  &    0.02  &   0.01  \\
2x3 & 1.31  & 0.05  & 0.02  & 0.01  \\
2x4 & 2.01  & 0.91  & 0.04  & 0.01  \\
2x5 & 3.70  & 2.45  & 0.12  & 0.05  \\
2x6 & 6.05  & 5.17  & 2.48  & 0.14  \\
3x3 & 3.21  &   0.11  &    0.04  &   0.01 \\
3x4 & 6.32  & 6.19  & 0.18  & 0.05 \\
3x5 & 10.40  & 17.66  & 0.48  & 0.22 \\
3x6 & 15.89  & 55.65  & 59.82  & 1.38 \\
4x4 & 7.12  &   15.49  &    8.41  &   0.18 \\
4x5 & 14.01  & 46.19  & 28.38  & 0.76 \\
4x6 & 19.81  & 67.20  & 86.95  & 35.32 \\
5x5 & 15.89  &   55.65  &   59.82  &   1.38 \\
5x6 & 41.52  & 139.59  & 182.42  & 110.65 \\
6x6 & 32.36  &   90.70  & 204.52  & 159.80 \\
\hline
\end{tabular}
\captionof{table}{$M\times N$ grids: optimal length of clearing search schedule and computation times in seconds.}
\label{fig0202010c}
\medskip
\end{minipage}

\subsection{Randomly Generated Trees \label{RandomBinaryTrees}}

In this section we apply the LVGS algorithm to trees in which children are
generated probabilistically. Specifically, we start with a complete binary
tree of depth one (one root with two children)\ and every vertex (except the
root)\ can have $n\in\left\{  0,1,2\right\}  $ children with probability
$p_{n}$ where $\left[  p_{0},p_{1},p_{2}\right]  =[\frac{1}{3},\frac{1}%
{3},\frac{1}{3}]$. We continue adding children to every leaf until the tree
reaches a maximum depth value $D_{\max}$. Hence a family of random binary
trees is fully described by the parameters $\mathbf{p}=[\frac{1}{3},\frac
{1}{3},\frac{1}{3}]$ and $D_{\max}$. In Table \ref{tabURandTrees5} we present,
for visibility range $L\in\{1,...,4\}$ and maximum depth $D_{\max}=5$, some
statistics averaged over $100$ randomly generated trees.

\begin{minipage}[c]{\textwidth}
\medskip
\centering
\begin{tabular}{|r|r|r|r|r|}
\hline
\multirow{2}{8em}{\textbf{max depth = 5}} & \multicolumn{4}{|c|}{\textbf{Visibility Range $L$}} \\ \cline{2-5}
& \textbf{1} & \textbf{2} & \textbf{3} & \textbf{4} \\
\hline
\hline
Number of Cleared Graphs & 81/100 & 100/100 & 100/100 & 100/100\\
Average Clearing Length & 5.32 & 3.19 & 1.32 & 0.36\\
Maximum Clearing Length & 14 & 14 & 6 & 2\\
Average Calculation Time & 11.99  & 22.75  & 31.62  & 18.90 \\
Maximum Calculation Time & 111.98  & 327.14  & 954.32 & 972.77\\
\hline
\end{tabular}
\captionof{table}{Statistics for 100 randomly generated trees with  $D_{max}=5$.}
\label{tabURandTrees5}
\medskip
\end{minipage}

\noindent In Table \ref{tabURandTrees6} we present similar results for 100
trees with maximum depth $D_{\max}=6$, and for visibility range $L\in
\{1,...,4\}$.

\begin{minipage}[c]{\textwidth}
\medskip
\centering
\begin{tabular}{|r|r|r|r|r|}
\hline
\multirow{2}{8em}{ \textbf{max depth = 6}} & \multicolumn{4}{|c|}{\textbf{Visibility Range $L$}} \\ \cline{2-5}
& \textbf{1} & \textbf{2} & \textbf{3} & \textbf{4} \\
\hline
\hline
Number of Cleared Graphs & 72/100 & 99/100 & 100/100 & 100/100\\
Average Clearing Length & 4.58 & 3.9 & 1.99 & 0.78\\
Maximum Clearing Length & 19 & 16 & 12 & 6\\
Average Calculation Time & 19.82  &  62.47  & 83.91  & 100.37 \\
Maximum Calculation Time &  303.04  &  1921.38 &  2467.45 &  4867.91\\
\hline
\end{tabular}
\captionof{table}{Statistics for 100 randomly generated trees with $D_{max}=6$.}
\label{tabURandTrees6}
\end{minipage}

\subsection{Grids with Randomly Deleted Edges\label{secNNDelGrids}}

In this section we use graphs which are created by deleting edges from a full
grid; for brevity, these will be referred as \textquotedblleft deleted
grids\textquotedblright. The specific method by which the edges are deleted is
the following. We start \ with a full grid and, for every edge $e$ of the grid:

\begin{enumerate}
\item If removal of $e$ would result in a disconnected graph, leave $e$ in place.

\item Otherwise remove $e$ with probability $p\in\left[  0,1\right]  $ (a parameter).
\end{enumerate}

\noindent The above procedure results in a deleted grid; higher $p$ values
reult in sparser graphs but, by construction, the resulting graph is always
connected. Results for $N\times N$ deleted grids are presented in the
following Tables \ref{fig0202011e},\ref{fig0202011c} and \ref{fig0202011d}.

\begin{minipage}[c]{\textwidth}
\medskip
\centering
\begin{tabular}{|r|r|r|r|r|}
\hline
\( \mathbf{p} = \mathbf{1/2}\) & \multicolumn{4}{|c|}{\textbf{Visibility Range $L$}} \\ \cline{2-5}
\textbf{grid size = 3x3} & \textbf{1} & \textbf{2} & \textbf{3} & \textbf{4} \\
\hline
\hline
Number of Cleared Graphs & 73/100 & 95/100 & 99/100 & 100/100\\
Average Clearing Length & 4.84 & 1.98 & 0.45 & 0\\
Maximum Clearing Length & 6 & 4 & 2 & 0\\
Minimum Clearing Length & 2 & 0 & 0 & 0\\
Average Calculation Time & 3.79  & 2.28  & 0.63  & 0.08 \\
Maximum Calculation Time & 8.62  & 5.87  & 2.38  & 0.23 \\
Minimum Calculation Time & 1.68  & 0.18  & 0.07  & 0.01 \\
\hline
\end{tabular}
\captionof{table}{Statistics for 100 $3\times 3$ deleted grids.}
\label{fig0202011e}
\medskip
\end{minipage}

\begin{minipage}[c]{\textwidth}
\medskip
\centering
\begin{tabular}{|r|r|r|r|r|}
\hline
\( \mathbf{p} = \mathbf{1/2}\) & \multicolumn{4}{|c|}{\textbf{Visibility Range $L$}} \\ \cline{2-5}
\textbf{grid size = 4x4} & \textbf{1} & \textbf{2} & \textbf{3} & \textbf{4} \\
\hline
\hline
Number of Cleared Graphs & 44/100 & 91/100 & 96/100 & 99/100\\
Average Clearing Length & 10.57 & 6.79 & 3.97 & 1.86\\
Maximum Clearing Length & 13 & 10 & 8 & 6\\
Minimum Clearing Length & 9 & 4 & 1 & 0\\
Average Calculation Time & 15.44  & 22.13  & 17.68  & 8.92 \\
Maximum Calculation Time & 42.02  & 62.15  & 46.21  & 21.42 \\
Minimum Calculation Time & 6.80  & 8.82  & 8.72  & 0.40 \\
\hline
\end{tabular}
\captionof{table}{Statistics for 100 $4\times 4$ deleted grids.}
\label{fig0202011c}
\medskip
\end{minipage}

\begin{minipage}[c]{\textwidth}
\medskip
\centering
\begin{tabular}{|r|r|r|r|r|}
\hline
\( \mathbf{p} = \mathbf{1/2}\) & \multicolumn{4}{|c|}{\textbf{Visibility Range $L$}} \\ \cline{2-5}
\textbf{grid size = 5x5} & \textbf{1} & \textbf{2} & \textbf{3} & \textbf{4} \\
\hline
\hline
Number of Cleared Graphs & 6/100 & 67/100 & 90/100 & 95/100\\
Average Clearing Length & 17.67 & 12.97 & 8.81 & 5.73\\
Maximum Clearing Length & 21 & 18 & 14 & 12\\
Minimum Clearing Length & 16 & 8 & 4 & 1\\
Average Calculation Time & 43.33  & 99.51  & 118.8  & 100.1 \\
Maximum Calculation Time & 118.19  & 282.3  & 438.73 & 378.94\\
Minimum Calculation Time & 18.30  & 37.70  & 43.56  & 41.53 \\
\hline
\end{tabular}
\captionof{table}{Statistics for 100 $5\times 5$ deleted grids.}
\label{fig0202011d}
\medskip
\end{minipage}

\section{Conclusion\label{sec05}}

We have formulated a graph search problem, with searchers looking for a mobile
target under the following assumptions: (a) the visibility field of the
searchers is limited, (b) movement is along the graph edges, (c) the searchers
have unit speed and (d) the target has infinite speed.

To solve the problem we have presented the LVGS\ algorithm, which is a
conversion the polygonal region search algorithm of
\cite{Guibas1997,Guibas1999}. This algorithm is guaranteed to always find a
solution if one exists. Furthermore, the LVGS algorithm\ can acommodate every
kind of visibility conditions (0-visibility, distance-based visibility of
arbitrary range and straight line visibility)\ using the visibility matrix.
While the problem formulation can acommodate an arbitrary number of searchers
and arbitrary searcher and target speed, the LVGS\ algorithm is particularly
suited for problems involving a single searcher, movement along the edges of
the graph, unit speed for the searcher and infinite speed for the target.

We conclude this paper by listing some research directions which we intend to
pursue in the future.

\begin{enumerate}
\item The polygonal region search algorithm of \cite{Guibas1997,Guibas1999} is
complex to implement and can be computationally impractical for complicated
regions. An alternative strategy to handle polygonal regions is to first
approximate them by discretization and then study the search problem on a
graph equivalent to the discretized region. We believe it will be interesting
to compare the performance of this approach combined with our LVGS\ algorithm
to that of the original polygonal region algorithm.

\item In the Introduction we have briefly mentioned the connection of graph
search to pursuit in graphs, especially to the problemof the cops and
invisible (or partially visible) robber. We believe that this connection
merits further study and that the LVGS\ algorithm could provide solutions
competitive to the ones obtained by combinatorial arguments
\cite{Clarke2019,Deren2011,Deren2012,Deren2013,Deren2015}.
\end{enumerate}

\end{document}